\documentclass[10pt]{article}
\usepackage{fullpage}
\usepackage{times}
\usepackage{amssymb}
\usepackage{amsmath}
 
\mathchardef\ordinarycolon\mathcode`\:
\mathcode`\:=\string"8000
\def\vcentcolon{\mathrel{\mathop\ordinarycolon}}
\begingroup \catcode`\:=\active
  \lowercase{\endgroup
  \let :\vcentcolon
  }

\newtheorem{theorem}{Theorem}
\newtheorem{lemma}{Lemma}

\newenvironment{proof}
  {\noindent \textbf{Proof:}}
  {$\blacksquare$}

\newcommand{\ket}[1]{|#1\rangle}
\newcommand{\bra}[1]{\langle#1|}
\newcommand{\ketbra}[2]{|#1\rangle\!\langle#2|}
\newcommand{\braket}[2]{\langle#1|#2\rangle}
\newcommand{\ox}{\otimes}
\newcommand{\smfrac}[2]{\mbox{$\frac{#1}{#2}$}}
\newcommand{\I}{\mathrm{i}}
\newcommand{\E}{\mathrm{e}}
\newcommand{\R}{\mathbb{R}}
\newcommand{\vnorm}[1]{|\!|#1|\!|_2}
\newcommand{\mnorm}[1]{|\!|\!|#1|\!|\!|_2}

\title{How Powerful is Adiabatic Quantum Computation?}
\author{Wim van Dam\thanks{Hewlett-Packard Laboratories, 
Mathematical Sciences Research Institute, 
Computer Science Division of the 
University of California at Berkeley (USA)
\texttt{vandam@cs.berkeley.edu}.
Supported by the \textsc{talent} grant S~62-552 of the Netherlands 
Organization for Scientific Research (NWO), 
the EU fifth framework project QAIP IST-1999-11234,
Darpa grant F30602-00-2-0601,
and an HP/MSRI post-doc fellowship.}
\qquad  
Michele Mosca\thanks{Centre for Applied Cryptographic Research,
Department of Combinatorics \& Optimization, 
University of Waterloo (Canada) \texttt{mmosca@uwaterloo.ca}}
\qquad 
Umesh Vazirani\thanks{Computer Science Division of the 
University of California at Berkeley (USA) \texttt{vazirani@cs.berkeley.edu}.
Supported in part by Darpa grant F30602-00-2-0601.}}

\begin{document}
\maketitle

\thispagestyle{empty}
\noindent {\textbf{Abstract:}\\ We analyze the computational power and limitations of 
the recently proposed `quantum adiabatic evolution
algorithm'.}

\section{Introduction}
Quantum computation is a revolutionary idea that has fundamentally
transformed our notion of feasible computation. The most dramatic 
example of the power of quantum algorithms was exhibited in Shor's celebrated quantum
algorithms for factoring and discrete log~\cite{shor}. Grover's quantum
search algorithm~\cite{grover} gives a quadratic speedup for a 
much wider class of
computational problems. Despite numerous attempts in the last few years,
it has proved to be a difficult challenge to design new quantum algorithms.
Recently, Farhi et al.\ \cite{farhi00,farhi01} proposed a novel paradigm for
the design of quantum algorithms --- via quantum adiabatic evolution. This
paradigm bears some resemblance to simulated annealing, in the sense that
the algorithm starts from an initial disordered state, and homes in on a 
solution (by what could be described as quantum local search) as a parameter
`$s$' is  smoothly varied from $0$ to $1$. The challenge
lies in showing that the process still converges to the desired 
solution with non-negligible probability if this transition is made 
in polynomial time.
In~\cite{farhi01,farhi3sat}, 
this paradigm was applied to the Exact Cover problem
(which has a close connection to the 3SAT problem), 
and using computer simulations it was shown that the algorithm works 
efficiently on small randomly chosen instances of this problem. 

In the first part of the article, we discuss the quantum adiabatic 
theorem and explain the quantum adiabatic approach to computation.
Next, we clarify the connection between the continuous time evolution of
adiabatic computing and the quantum circuit model with its 
discretized time.
We do this by describing a way of efficiently simulating quantum 
adiabatic algorithms with a network of standard quantum gates.
After this exposition, we explore 
three questions about quantum adiabatic evolution algorithms.

Can we apply the exponential lower bounds for quantum search~\cite{bbbv} to 
conclude that the adiabatic quantum algorithm for 3SAT must take exponential time? 
More concretely, at a high level of abstraction, the adiabatic
 quantum algorithm
for 3SAT may be viewed as some quantum process that gets information about the
3SAT instance only by (quantum) queries of the following type: given a truth
assignment, how many clauses of the formula $\Phi$ are not satisfied? 
We prove that there is a (classical) polynomial time algorithm that 
can reconstruct the 3CNF formula $\Phi$
by making polynomially many queries of this type. It is somewhat surprising
that this question does not appear to have been studied in the context of
relativization results for NP. In our context, it rules out any query 
complexity based (quantum) lower bound for the adiabatic quantum solution 
of 3SAT.

Is adiabatic quantum computing really quantum? We give an example of an
adiabatic quantum algorithm for searching that matches the optimal quadratic 
speedup obtained by Grover's search algorithm. This example demonstrates 
that the `quantum local search', which is implicit in the adiabatic evolution,
 is truly non-classical in nature from a computational viewpoint. 

Finally, we give a simple example of a computational problem on which
the adiabatic quantum algorithm provably takes exponential time. Although
the problem is easy to solve classically, it is designed to be difficult
for algorithms based on local search: its global optimum lies in a
narrow basin, while there is a local optimum with a much larger
basin. Let $f$ be a function on the $n$-bit strings, where
$f(x)$ depends only on $w(x)$, the Hamming weight of $x$. The problem is to find
an $x$ that minimizes $f(x)$. (Obviously, it is straightforward to solve this
class of problems in $n+1$ steps.) Consider functions $f$ such that 
for $w(x) \leq (\smfrac{1}{2} + \varepsilon)n$, $f(x) = w(x)$, and which decreases
for $w(x)>(\smfrac{1}{2}+\varepsilon)n$ to the global minimum 
$f(1^n) = -1$. We prove that for such instances,
the adiabatic quantum algorithm requires an exponential slowdown in
$n$. We do this by showing that the gap between the minimum and second 
eigenvalue of the Hamiltonian of the system is exponentially small. 
In an upcoming paper~\cite{dv}, we generalize these techniques to show a similar
exponential slowdown for 3SAT.

\section{The Quantum Adiabatic Theorem}
The Hamiltonian of a physical system gives a complete 
specification of the time evolution of this system.
At a given time $t$, let $\psi(t)$ denote the state of
the system 
under the influence of the Hamiltonian $H(t)$. 
The differential equation that describes the time evolution
 is the well-known Schr{\"o}dinger equation:
\begin{eqnarray*}
\I\hbar\smfrac{d}{dt}\ket{\psi(t)} & = & H(t)\ket{\psi(t)},
\end{eqnarray*}
where $\hbar$ is Planck's constant $h\approx 6.63\times 10^{-34}$ 
Joule-second, divided by $2\pi$. 
A Hamiltonian is described by a Hermitian matrix, 
whose eigenvectors represent the eigenstates of the
system.  The corresponding eigenvalues refer to the 
different energies of the eigenstates.
The state (eigenvector) with the lowest energy (eigenvalue) 
is called the `ground state' of the system.
The Schr\"odinger equation can also be described 
with reference to the unitary transformation $U$
that is defined by the Hamiltonian $H(t)$
(from now on we work with $\hbar=1$):
\begin{eqnarray*}
\smfrac{d}{dt}U(t) & = & -\I H(t) U(t),
\end{eqnarray*}
 with the initial condition $U(0)=I$.
We say that the Hamiltonian evolution 
from $H(0)$ to $H(T)$ \emph{induces} 
the unitary transformation $U(T)$. 
The evolution of a system with a time-independent 
Hamiltonian $H$ is easily expressed by the exponential
$U(T)=\E^{-\I T H}$.
Finding the (approximate) solutions for Hamiltonians
that vary in time is one of the core tasks in
quantum physics. 
One of the most important cases of such a 
time-dependent case is described by the adiabatic 
evolution of an isolated quantum mechanical system.

The quantum adiabatic theorem states that a 
physical system that is initially in its ground state,
 tends to stay in this lowest energy state, provided that 
the Hamiltonian of the system is changed `slowly 
enough'.\cite{bornfock}

The quantitative version of the adiabatic theorem
gives the following
specific upper bound on the slowdown that is required for the
adiabatic evolution of the ground state.
(See for example \cite{messiah} for more details on this.)
Parameterize the time-dependent Hamiltonian by 
$H(s)$ for $0\leq s \leq 1$ and its
ground state by $\phi(s)$.  Our goal is thus
to gradually transform the applied Hamiltonian from $H(0)$ to $H(1)$ 
such that the initial state $\psi(0)=\phi(0)$ evolves to a close approximation
 $\psi(1)\approx\phi(1)$ of the ground state of $H(1)$.  
We introduce a delay factor $\tau(s)$, which determines the rate at 
which the Hamiltonian is modified as a function of $s$. Now the 
Schr{\"o}dinger equation in $s$ equals
\begin{eqnarray*}
\smfrac{d}{ds}\ket{\psi(s)} & = & -\I \tau(s) H(s)\ket{\psi(s)}.
\end{eqnarray*}
The crucial quantity for this transformation is the 
gap between the two smallest eigenvalues of $H(s)$,
which we denote by $g(s)$. It can be shown that a 
delay schedule $\tau$ with 
\begin{eqnarray*}
\tau(s) & \gg & 
\frac{\mnorm{\smfrac{d}{ds}H(s)}}
{g(s)^{2}}
\end{eqnarray*}
is `sufficiently slow' for the adiabatic evolution
from $\phi(0)$ to $\phi(1)$.
As a result, the total delay of this process will be 
of the order $\int_{s=0}^1{\tau(s)ds}$. 
For most Hamiltonians it is too difficult to determine
the gap $g(s)$ for every $s$.  
If this is the case, we can also look at the \emph{minimum gap}
$g_\mathrm{min} := \min_s {g(s)}$ and the maximum 
$\Delta_\mathrm{max} := 
\max_s {\mnorm{{\smfrac{d}{ds}}H(s)}}$, 
and obtain the adiabatic evolution with the constant delay factor
$\tau(s) = \tau_c \in O(\smfrac{\Delta_\mathrm{max}}{g_\mathrm{min}^2})$.

\section{Adiabatic Quantum Computation}
\emph{Adiabatic quantum computation,} as proposed
by Farhi et al.\cite{farhi00}, works as follows.
At time $t=0$, the quantum mechanical system is
described by a Hamiltonian $H_0$, whose eigenstates are 
easy to compute.  Next, this system is slowly transformed
to its final Hamiltonian $H_f$, for which the ground state
is the solution to a specific minimization problem $f$. 
We do this is by letting the energies $\lambda_z$ of the 
eigenstates $z$ of $H_f$ correspond with the 
function that we try to minimize. 
Hence, if this function $f$ has domain $\{0,1\}^n$,
then the final Hamiltonian is defined by 
\begin{eqnarray*}
H_f & := & \sum_{z\in\{0,1\}^n}{f(z)\cdot\ketbra{z}{z}}.
\end{eqnarray*}
We will assume throughout this paper that $f:\{0,1\}^* \rightarrow \R$ is 
computable in polynomial time, and that $f(x)$ is bounded by a 
polynomial in $|x|$.

The choice of the initial Hamiltonian $H_0$ is
independent of the solution of the problem,
and will be such that $H_0$ is not diagonal 
in the computational $z$-basis.  
Specifically, we consider the `Hadamard basis' with 
the bit values
\begin{eqnarray*}
\ket{\hat{0}} := \smfrac{1}{\sqrt{2}}(\ket{0}+\ket{1})
& \mbox{ and } &
\ket{\hat{1}} := \smfrac{1}{\sqrt{2}}(\ket{0}-\ket{1}).
\end{eqnarray*}
For a binary string $z\in\{0,1\}^n$, let $\ket{\hat{z}}$
denote the state which would be written as $\ket{z}$ in this
 basis.
(The unitary mapping between these two representations is
provided by the $n$-fold Hadamard matrix: 
$W^{\ox n}\ket{z}= \ket{\hat{z}}$ 
and $W^{\ox n}\ket{\hat{z}}= \ket{z}$.)
 
A simple starting Hamiltonian that fulfills the above 
requirements  is 
\begin{eqnarray*}
H_0 & := & \sum_{z\in\{0,1\}^n}{h(z)\cdot\ketbra{\hat{z}}{\hat{z}}},
\end{eqnarray*} 
with $h(0^n)=0$ and $h(z)\geq 1$ for all other $z\neq 0^n$,
such that the ground state with zero energy of $H_0$ is
 the uniform superposition 
$\ket{\hat{0}\cdots\hat{0}}=\smfrac{1}{\sqrt{2^n}}\sum_z{\ket{z}}$.
Having defined the initial and final conditions of 
our system, we will now describe the time-evolution.

Following the proposal by Farhi et al.\ in \cite{farhi00,farhi01},
we can define the time dependent Hamiltonian $H(t)$ as 
the linear combination of the starting and the final Hamiltonian:
\begin{eqnarray*}
H(t) & := & \left(1-\smfrac{t}{T}\right)H_0 + \smfrac{t}{T} H_f,
\end{eqnarray*}
with $0\leq t \leq T$, and $T$ the crucial \emph{delay factor} of the 
$H_0\rightarrow H_f$ transition.

By the adiabatic theorem we know that this system will map
the initial ground state $\ket{\psi(0)}=\ket{\hat{0}^n}$ to the global 
minimum of the function $f$, provided that we pick $T$ large enough.
In the previous section we mentioned that 
$T\in O({\Delta_\mathrm{max}}{g^{-2}_\mathrm{min}})$ is a sufficient 
upper bound on this delay.  
Without any further knowledge about the specific Hamiltonian 
$H(t)$ --- which involves detailed knowledge about the function $f$,
this is also a lower bound for a reliable
adiabatic evolution from $H_0$ to $H_f$.
Because $\mnorm{\smfrac{d}{ds}H(s)}$ is 
 polynomial in $n$ (as long as $f\in\mathrm{poly}(n)$),
we will ignore this factor and focus mostly
on the $T \gg {g^{-2}_\mathrm{min}}$ requirement
for the delay of the adiabatic quantum computation.

\section{Approximating the Adiabatic Evolution}
In this section we explain how the 
continuous time evolution from $H_0$ to $H_f$
can be approximated by a quantum circuit of size
$\mathrm{poly}(nT)$. Our goal is to demonstrate the
ingredients of the polynomial upper bound, and we 
do not try to optimize to get the most efficient 
simulation.

The approximation is established in two steps.
First, we discretize the evolution from $H_0$ to
$H_f$ by a finite sequence of Hamiltonians
$H'_1$, $H'_2,\dots$ that 
gives rise to the same overall behavior.
Second, we show how at any moment the combined 
Hamiltonian $H'_j = (1-s)H_0 + s H_f$ can be 
approximated by interleaving two simple 
unitary transformations.

To express the error of our approximation, we 
use the $\ell_2$ induced operator norm ``$\mnorm{\circ}$'':
\begin{eqnarray*}
\mnorm{M} & := & \max_{\vnorm{x}=1}{\vnorm{M x}}.
\end{eqnarray*}
The next lemma compares two Hamiltonians 
$H(t)$ and $H'(t)$ and their respective 
unitary transformations $U(T)$ and $U'(T)$.
\begin{lemma}
Let $H(t)$ and $H'(t)$ be two time-dependent 
Hamiltonians for $0\leq t \leq T$, and 
let $U(T)$ and $U'(T)$ be the respective 
unitary evolutions that they induce. 
If the difference between the Hamiltonians is 
limited by $\mnorm{H(t)-H'(t)}\leq \delta$ for every $t$, 
then the distance between the induced transformations  
is bounded by $\mnorm{U(T)-U'(T)}\leq \sqrt{2T\delta}$.
\end{lemma}
\begin{proof}
Let $\psi(t)$ and $\psi'(t)$ be the two state trajectories  
of the two Hamiltonians $H$ and $H'$ with initially
$\psi(0)=\psi'(0)$. 
Then, for the inner product between the two states
(with initially $\braket{\psi'(0)}{\psi(0)}=1$), 
we have
\begin{eqnarray*}
\smfrac{d}{dt}
{\braket{\psi'(t)}{\psi(t)}} & = & 
-\I\bra{\psi'(t)}(H(t)-H'(t))\ket{\psi(t)}.
\end{eqnarray*}  
Because at any moment $t$ we have 
$\vnorm{\ket{\psi(t)}}=\vnorm{\ket{\psi'(t)}}=1$ 
and $\mnorm{H(t)-H'(t)}\leq \delta$, we see 
that at $t=T$ the lower bound
$|\braket{\psi'(T)}{\psi(T)}|\geq 1-T\delta$
holds.
This confirms that for every vector $\psi$ we have
$\vnorm{U(T)\ket{\psi}-U'(T)\ket{\psi}}\leq \sqrt{2T\delta}$.
\end{proof}

This lemma tells us how we can deviate from the ideal 
Hamiltonian $H(t):=(1-\smfrac{t}{T})H_0 + \smfrac{t}{T}H_f$, 
without introducing too big of an error to the induced evolution. 
As mentioned above, we will approximate the continuous 
$H(0)\rightarrow H(T)$ trajectory by a sequence 
of $r$ Hamiltonians $H'_1,\dots, H'_r$, each of which
applied for a duration of $\smfrac{T}{r}$.
This yields the unitary evolution $U'(T)$, defined by
\begin{eqnarray*}
U'(T) & := & \E^{-\I(\smfrac{T}{r})H'_r}
\cdots\E^{-\I(\smfrac{T}{r})H'_1},
\end{eqnarray*}
with for any $1\leq j \leq r$ the Hamiltonian
$H'_j := H(\smfrac{jT}{r}) = 
(1-\smfrac{j}{r})H_0 + (\smfrac{j}{r})H_f$.
If we view $H'$ as a time-dependent Hamiltonian 
$H'(t) := H_{j(t)}$ with $j(t) := {\lceil\smfrac{rt}{T}\rceil}$, 
then we have the bound 
$\mnorm{H(t)-H'(t)} \leq \smfrac{1}{r}\mnorm{H_f-H_0} \in 
O(n^d/r)$ for all $t$.  By the previous lemma 
we thus have the bound $\mnorm{U(T)-U'(T)} \in O(\sqrt{Tn^{d}/r})$.

The second part of our approximation deals with the problem 
of implementing the unitary transformations $U'_j$ defined by 
\begin{eqnarray*}
U'_j & := & \E^{-\I \frac{T}{r} (1-\frac{j}{r})H_0 -\I
\frac{T}{r}(\frac{j}{r})H_f}.
\end{eqnarray*}
with elementary operations.

The Campbell-Baker-Hausdorff theorem\cite{bhatia} tells us how
well we can approximate `parallel Hamiltonians' by 
consecutive ones:
$\mnorm{\E^{A+B}-\E^A\E^B} \in O(\mnorm{AB})$.
Hence in our case, by defining
\begin{eqnarray*}
U''_j & := & \E^{-\I \frac{T}{r}(1-\frac{j}{r})H_0}
\cdot\E^{-\I\frac{T}{r}(\frac{j}{r})H_f},
\end{eqnarray*}
we get the approximation $\mnorm{U'_j-U''_j}
\in O(\smfrac{T^2}{r^2}\mnorm{H_0 H_f})$.
This leads to 
$\mnorm{U'(T)-U''(T)}\in O(n^{d+1}T^2/r)$, and
hence also for the original transformation:
$\mnorm{U(T)-U''(T)}\in O(n^{d+1}T^2/r)$.

Because $H_0 = \sum_z{h(z)\ketbra{\hat{z}}{\hat{z}}}$ 
is diagonal in the Hadamard basis $\{\hat{0},\hat{1}\}^n$,
and $H_f = \sum_z{f(z)\ketbra{z}{z}}$ is diagonal in the
computational bases, we can implement 
the above $U''_j$ as
\begin{eqnarray*}
U''_j & = & W^{\ox n} \cdot F_{0,j} \cdot W^{\ox n} \cdot F_{f,j},
\end{eqnarray*}
with $W^{\ox n}$ the $n$-fold Hadamard transform, and $F_0$ and $F_f$
the appropriate phase changing operations:
\begin{eqnarray*}
F_{0,j}\ket{z} & :=  & \E^{-\I \frac{T}{r}(1-\frac{j}{r})h(z)}\ket{z}, \\
F_{f,j}\ket{z} & := & \E^{-\I\frac{T}{r}(\frac{j}{r})f(z)}\ket{z}.
\end{eqnarray*}
Because $h(z)$ and $f(z)$ are easy to compute, so are 
$F_0$ and $F_f$. 
We have thus obtained the following theorem.

\begin{theorem}
Let $H_0$ and $H_f$ be the initial and final Hamiltonians used
in an adiabatic computation, with the function $f\in O(n^d)$. 
Then, the unitary transformation $U(T)$ induced by the 
time-dependent Hamiltonian $H(t) := (1-\smfrac{t}{T})H_0 + 
\smfrac{t}{T}H_f$ can be approximated by $r$ consecutive 
unitary transformations $U''_1,\dots,U''_r$
with $r\in O(T^2n^{d+1})$.
Furthermore, each $U''_j$ has the form $W^{\ox n} F_0 W^{\ox n} F_f$
and can thus be efficiently implemented in $\mathrm{poly}(nT)$ time.
\end{theorem}
It is interesting to note that the $W^{\ox n} F_0 W^{\ox n} F_f$
transformation has the same form as the `Grover iteration' of
the standard quantum search algorithm\cite{grover}.  
More recently, we also learned that the work of Hogg on quantum search
heuristics\cite{hogg} describes essentially the same algorithm as the 
adiabatic approach to minimization.

\section{Quantum Adiabatic Searching}
One question that should be asked first is if
adiabatic quantum computing is truly quantum 
computing.  
In this section we answer this question affirmatively 
by reproducing the quadratic speed-up of Lov Grover's 
search algorithm. 

For the search problem, the function 
$f:\{0,1\}^n \rightarrow \R$
takes on value $1$ on all strings except the solution $u \in \{0,1\}^n$
for which $f(u) = 0$. Thus the final Hamiltonian for the 
adiabatic algorithm, $H_u$, will have eigenstates
$\ket{z}$ with eigenvalue $1$, with the exception of 
the unknown solution $u \in \{0,1\}^n$, which has 
eigenvalue $0$:
\begin{eqnarray*}
H_u & := & \sum_{z\in\{0,1\}^n\setminus \{u\}}{\ketbra{z}{z}}.
\end{eqnarray*}

The initial Hamiltonian is defined similarly, except that
it is diagonal in the Hadamard basis, and has ground state $\ket{\hat{0}^n}$: 
\begin{eqnarray*}
H_0 & := & \sum_{z \in\{0,1\}^n\setminus\{0^n\}}{\ketbra{\hat{z}}{\hat{z}}}.
\end{eqnarray*}

With these initial and final conditions one can easily show
that for the resulting time-dependent Hamiltonian
\begin{eqnarray*}
H(t) & := & (1-\smfrac{t}{T})H_0 + \smfrac{t}{T}H_u ,
\end{eqnarray*}
the gap between the two smallest eigenvalues as a function
of $s :=\smfrac{t}{T}$ is expressed by 
\begin{eqnarray}\label{eq:gminsearch}
g(s) & = & 
\sqrt{\frac{2^n+4(2^n-1)(s^2-s)}{2^n}}.
\end{eqnarray}
This gap reaches its minimum at $t=\smfrac{T}{2}$
when it equals $\smfrac{1}{\sqrt{2^n}}$.
At first sight, this would lead to the conclusion that
the necessary delay factor $T=\Omega(g_{\mathrm{min}}^{-2})$ is 
linear in $N = 2^n$.  
However, by using our knowledge of the gap
function $g(\smfrac{t}{T})$ 
 we can significantly reduce the running time to $O(\sqrt{N})$.

For example, regardless of the solution $u$, we know that 
the transition from $H(0)$ to $H(\smfrac{T}{3})$ will have
 a minimal gap 
that is significantly bigger than $\smfrac{1}{\sqrt{N}}$.
The necessary delay factor that we use for this first part 
of our transformation $H_0\rightarrow H_u$, can therefore 
be much smaller than $N$.
In general at any moment $s=\smfrac{t}{T}$, 
Equation~\ref{eq:gminsearch} tells us the size of the gap 
$g(s)$, and hence
the delay factor that suffices at that moment.
This means that we can employ a varying delay factor 
$g(s)^{-2}$, 
without destroying the desired adiabatic properties of the 
evolution $H_0\rightarrow H_u$.
In sum, this approach leads to a total delay factor of
\begin{eqnarray*} 
\int_{s=0}^{1}{\frac{ds}{g(s)^2}} & = & 
\int_{s=0}^{1}{\frac{2^n}{2^n+4(2^n-1)(s^2-s)} ds}
\\
& = & 
\frac{2^n\cdot\arctan(\sqrt{2^n-1})}{\sqrt{2^n-1}}.
\end{eqnarray*}
As a function of $N=2^n$, this gives a time 
complexity $O(\sqrt{2^n}) = O(\sqrt{N})$,
which coincides with the well-known square root speed-up
of quantum searching.  
(See the article by Farhi and Gutmann\cite{farhisearch}
 for another example of a `continuous time algorithm' 
for quantum searching.)

\section{Query Bounds for the 3SAT Problem}

The adiabatic quantum algorithms of \cite{farhi01, farhi3sat}
work on 3SAT as follows: on input a formula 
$\Phi = C_1 \wedge \dots \wedge C_M$ (where the $C_i$ are clauses 
in variables $x_1, \dots, x_n$), the only way the quantum algorithm
gathers information about $\Phi$ is by queries which ask, for a 
given truth assignment $b$ (in general a superposition of assignments), 
how many of the $M$ clauses $b$ does not satisfy. 
A natural approach to establishing a lower bound on the running time 
of the adiabatic quantum algorithm is to show that any quantum algorithm
must make a large number of such queries to solve the problem.
This is the approach that
leads to the exponential lower bound for unstructured search~\cite{bbbv}
(there the query asked, for a given assignment $b$, whether or not it
is a satisfying assignment),
thus showing that relative to a random oracle NP is not a subset of 
subexponential quantum time. In this section, we show that the seemingly
small difference between the specifications of these two types of queries
results in a dramatic change in the query complexity --- 
$O(n^3)$ queries suffice to obtain enough information to characterize $\Phi$.
Thus black box or oracle techniques do not rule out a polynomial time
solution to 3SAT by adiabatic quantum search. To reconcile this with 
the oracle results from~\cite{bbbv}, it is useful to recall
that the Cook-Levin theorem, suitably formulated as saying that 
NP has a `local-checkability' property, does not relativize~\cite{AIV}
(see \cite{Va} for a brief discussion of this issue). 
In this sense, the results in this section indicate that even keeping
track about the number of unsatisfied clauses constitutes sufficient structural
information about the problem to bypass the oracle results. 

More formally, let
\begin{eqnarray*}
F_{\Phi}(b) & := & \mbox{``\# unsatisfied clauses in
assignment $\Phi(b)$''},
\end{eqnarray*}
with $b\in\{0,1\}^n$. In our black box model, the quantum algorithm
is only allowed to 
access $\Phi$ via a quantum black-box $B_\Phi$ that reversibly maps
$\ket{b}
\ket{0} \mapsto \ket{b}
\ket{F_{\Phi}(b)}$. In this section, we prove that the 
query complexity for 3SAT is $O(n^3)$, by showing that
$F_{\Phi}$ is completely determined by its values on the 
$O(n^3)$ input strings of Hamming weight $\leq 3$.
Our techniques also apply to the
Exact Cover problem discussed in \cite{farhi01}.

For convenience, and without loss of generality, we will not allow
repeated variables in the same clause, but instead will allow
clauses of size less than $3$.  For example, we can replace the
clause $(x_1 \vee x_1 \vee x_2)$ with $(x_1 \vee x_2)$, and $(x_1
\vee \neg x_1 \vee x_2)$ with a constant clause $(1)$ that
is always satisfied.  Without loss of generality, we can assume
that the number of such $(1)$ clauses is $0$.

Let us introduce some notation. Let $|XXX|$ denote the number of
clauses in $\Phi$ that have all three variables without negation
(e.g. $(x_1 \vee x_2 \vee x_3)$). We will say that these clauses
are ``of the form'' $XXX$. Let $|\overline{X}XX|$ denote the number
of clauses that have exactly one variable negated (e.g.\
$(x_1 \vee \neg x_2 \vee x_3)$). Further, we let
$|\overline{X}\overline{X}X|$ denote the number
of clauses that have exactly two variable negated, and
$|\overline{X}\overline{X}\overline{X}|$ denote the number of
clauses that have all three variables negated. We also define the
analogous $1$ and $2$ variable versions of these expressions.

Furthermore, if we subscript any of the $X$ with an index, say $i$,
then we only count clauses that have $x_i$ as one of the
non-negated (or
\emph{positive}) variables. Similarly, if we subscript any of the
$\overline{X}$ with an index, say $i$, then we only count clauses
that have $\overline{x_i}$ as one of the negated variables. For
example, $|X_iXX|$  denotes the number of clauses in $\Phi$ that
contain the variable $x_i$ and two other positive variables,
$|\overline{X}_i X_j X |$ denotes the number of clauses with
$\overline{x_i}$ and $x_j$ and another positive variable, and $|X_i
X_j X |$ denotes the number of clauses that have one of the
positive variables equal to $x_i$, another equal to $x_j$, and
another positive variable. The expression $|\overline{X}_i
\overline{X}_j X_k|$ equals the number of times the clause
$( \neg x_i \vee \neg x_j \vee x_k)$ (or equivalent permuted
clauses like $( x_k \vee \neg x_j \vee \neg x_i)$)  occurs.

These expressions are symmetric under permutation of the symbols,
so for example, $|X_i X_j X |=|X_j X_i X|$ and $|\overline{X}_i X_j
X_k | = |\overline{X}_i X_k X_j |$.

For example, we have that
\begin{eqnarray*}
F_\Phi(0^n) &=& |XXX| + |XX| + |X|
\end{eqnarray*}
since any clause with a negated variable will be satisfied, and the
rest will not be satisfied.

The following definitions will be helpful. For each $i \in
\{1,\dots, n\}$ let
\begin{eqnarray*}
 Y_i & := &
|\overline{X_i}XX| - |X_i XX| + |\overline{X_i}X| - |X_i X| +|\overline{X_i}| - |X_i| .
\end{eqnarray*}
For each pair $i,j \in \{1,\dots,n\}$, $i \neq j$, let
 \begin{eqnarray*}
 Y_{ij} & := &
|\overline{X_i}\overline{X_j}X| + |X_i X_j X| -
|\overline{X_j}X_iX| - |\overline{X_i}X_jX|
 + |\overline{X_i}\overline{X_j}| + |X_i X_j|
-|\overline{X_i}X_j| - |\overline{X_j}X_i|.
 \end{eqnarray*}
 For each triple $i,j,k$ of pairwise distinct integers from
$\{1,\dots,n\}$,  let
 \begin{eqnarray*}
 Y_{ijk} & := &
 |\overline{X_k}X_i X_j| + |\overline{X_i}X_jX_k| + |\overline{X_j}X_i X_k|
  +|\overline{X_i}\overline{X_j}\overline{X_k}|  - |X_i X_j X_k|
 -|\overline{X_j}\overline{X_i}X_k| -
|\overline{X_i}\overline{X_k}X_j| -
|\overline{X_j}\overline{X_k}X_i| .
 \end{eqnarray*}

For each $i \in \{1,\dots, n\}$ let $e^i$ denote the
string with a $1$ in the $i$th position and $0$s elsewhere. For
each $i,j \in \{1,\dots, n\}$, $i \neq j$, let $e^{ij}$
denote the string with a $1$ in positions $i$ and $j$ and $0$s
elsewhere.  For each $i,j,k \in \{1,\dots, n\}$, pairwise
distinct, let $e^{ijk}$ denote the string with a $1$ in
positions $i,j$ and $k$ and $0$s elsewhere.

We now have the next theorem.

\begin{theorem}
Let $b \in \{0,1\}^n$ and let $I$ be the subset of
$\{1,\dots, n\}$ such that $b_i = 1 \iff i \in I$. Then
\begin{eqnarray*}
F_\Phi(b) &=& F_\Phi(0^n) + \sum_{i \in I} Y_i + \sum_{i < j
\in I} Y_{ij} + \sum_{i<j<k \in I} Y_{ijk}.
\end{eqnarray*}
Furthermore,
\begin{eqnarray*}
 Y_i &=& F_\Phi(e^i) - F_\Phi(0^n) \\
Y_{ij}& =& F_\Phi(e^{ij}) - F_\Phi(e^i) - F_\Phi(e^j) +
F_\Phi(0^n) \\
 Y_{ijk} &=& F_\Phi(e^{ijk})
- F_\Phi(e^{ij}) - F_\Phi(e^{ik})-F_\Phi(e^{jk}) 
  + F_\Phi(e^i) + F_\Phi(e^j) + F_\Phi(e^k)
- F_\Phi(0^n).
\end{eqnarray*}
\end{theorem}
In other words, in order to be able to evaluate $F_\Phi$
for every input string $\{0,1\}^n$, we only need to query
the black-box $B_\Phi$ on the $O(n^3)$ inputs with Hamming weight at most
$3$
(the cases $b\in\{0^n,e^{i},e^{ij},e^{ijk}\}$).
Specifically, we can decide whether $\Phi$ is satisfiable or not
by querying the black-box $B_\Phi$ a total of $O(n^3)$ times,
after which we use the query results to evaluate
$F_\Phi$ for all other possible inputs  $b\in\{0,1\}^n$.
 If any of the strings give $F_\Phi(b)=0$, then $\Phi$ is satisfiable,
 otherwise it is not satisfiable.
(Clearly, with this information we can also answer other decision
problems like ``$\Phi\in \mathrm{PP}$?'')
The full proof of this theorem is described in the appendix
of this article.

\section{Lower Bounds for Adiabatic Algorithms}
In this section we present an easy $n$-bit problem,
for which the adiabatic approach only succeeds 
if it is allowed an exponential delay.
We do this by changing an easy problem 
(the Minimum Hamming Weight Problem)
into a perturbed version for which the 
proper solution is as far as possible from 
its local minimum.  
It will be shown that for this perturbed version,
the quantum adiabatic algorithm does indeed 
require exponential time.

\subsection{The Minimum Hamming Weight Problem}
Consider the adiabatic quantum algorithm that
tries to minimize the Hamming weight $w(z)$ of an
$n$ bit string $z\in\{0,1\}^n$.
We define the initial Hamiltonian by 
$H_0 := \sum_{z} w(z)\ketbra{\hat{z}}{\hat{z}}$, 
such that the time-dependent Hamiltonian is
\begin{eqnarray*}
H_w(t) & := & 
\left(1 - \smfrac{t}{T}\right)
\sum_{z\in\{0,1\}^n}{ w(z)\ketbra{\hat{z}}{\hat{z}}}
  +
\smfrac{t}{T}\sum_{z\in\{0,1\}^n}{ w(z)\ketbra{z}{z}}. 
\end{eqnarray*}
As intended, the ground state of the final Hamiltonian 
is simply $\ket{0\cdots 0}$ with zero energy. 

Since $w(z) = z_1 + \cdots + z_n$, it is easy to see that 
$H_w(t)$ is a sum of $n$ Hamiltonians, each acting on a 
single qubit. Thus even though $H_w(t)$ is a 
$2^n\times 2^n$ dimensional
matrix, which thus has $2^n$ eigenstates, these eigenstates 
and their corresponding eigenvalues may be computed by solving
the $2$ dimensional problem. 
For the analysis of the minimal gap between the two smallest
eigenvalues it is again convenient to
introduce a relative time-parameter $s := \smfrac{t}{T}$,
which ranges from $0$ to $1$. The eigen-decomposition for
the $2$ dimensional problem yields:
\begin{eqnarray} \label{eq:eigenw}
\frac{1}{2}
\left(\begin{array}{rr}
1-s & s-1 \\ s-1 & 1+s
\end{array}\right) & = &
\begin{array}{l}
\lambda_0(s)\ketbra{v_0(s)}{v_0(s)} 
+\lambda_1(s)\ketbra{v_1(s)}{v_1(s)}.
\end{array}
\end{eqnarray}
with
\begin{eqnarray*}
\lambda_0(s) =  \smfrac{1}{2} -
\smfrac{1}{2}\sqrt{2s^2-2s+1} &\mbox{ and }& 
\lambda_1(s) =  \smfrac{1}{2} +
\smfrac{1}{2}\sqrt{2s^2-2s+1}.
\end{eqnarray*}
Specifically, at $s=0$ we have
$\ket{v_0(0)} = \ket{\hat{0}} = \frac{1}{\sqrt{2}}(\ket{0}+\ket{1})$
and
$\ket{v_1(0)} = \ket{\hat{1}} = \frac{1}{\sqrt{2}}(\ket{0}-\ket{1})$,
while at $s=1$
we have $\ket{v_0(1)} = \ket{0}$ and $\ket{v_1(1)} = \ket{1}$.

For the $n$ qubit case, it is easily shown that for every $y\in\{0,1\}^n$ 
there is an eigenvalue
\begin{eqnarray*}
\lambda_y(s) & = & (n-w(y))\cdot\lambda_0(s) + w(y)\cdot\lambda_1(s),
\end{eqnarray*}
where the corresponding eigenvector is the $n$-fold tensor product
\begin{eqnarray*}
\ket{v_y(s)} & := & 
\ket{v_{y_1}(s)} \ox \ket{v_{y_2}(s)} \ox \cdots \ox 
\ket{v_{y_n}(s)}.
\end{eqnarray*}

Because $\lambda_0(s) < \lambda_1(s)$ for all $s$,
 the ground state of $H(s T)$
is $\ket{v_0(s),\dots,v_0(s)}$ with eigenvalue $n\lambda_0(s)$.
The eigenvalues closest to this ground energy 
are those associated with the $w(y)=1$ eigenvectors $\ket{v_y(s)}$,
which have eigenvalue $(n-1)\lambda_0(s) + \lambda_1(s)$. 
Hence, the energy gap  between the two smallest eigenvalues 
is $g(s) = \sqrt{2s^2-2s+1}$, 
with its minimum $g_{\mathrm{min}} = \frac{1}{\sqrt{2}}$
 at $s=\smfrac{1}{2}$ ($t=\smfrac{T}{2}$).
Because this gap is independent of $n$, we can
transform $H_0$ to $H_w$ adiabatically with a 
constant delay factor.
As a result, the ground state 
$\ket{v_0^n(s)} := \ket{v_0(s)}^{\ox n}$ 
of the system evolves from $\ket{\hat{0}\cdots \hat{0}}$ to 
$\ket{0 \cdots 0}$ in time $O(1)$. 

We will now discuss an important aspect of the above 
adiabatic evolution, which we will use in the lower bound of 
the next section.  We saw how the initial ground state 
of the Hamiltonian $H_0$ is the uniform superposition
$\smfrac{1}{\sqrt{2^n}}\sum_z{\ket{z}}$ while the
final ground state of $H_w$ is the zero string  
$\ket{0^n}$. Both states share the property that they have an
exponentially small component in the 
subspace spanned by computational basis vectors 
labeled  with strings of Hamming weight at least $(\smfrac{1}{2} + \varepsilon)n$.
With the eigenvector decomposition of Equation~\ref{eq:eigenw} 
we can see that such an upper bound holds for $0\leq s\leq 1$.
Take for example the vector $\ket{1^n}$, which indeed has:
\begin{eqnarray} \label{eq:lowerbw}
|\braket{1^n}{v_0(s)\cdots v_0(s)}| & \leq & \smfrac{1}{\sqrt{2^n}},
\end{eqnarray}
for all $s$.
This bound suggests that a perturbation of the Hamiltonian $H_w$ in 
this subspace
will only have an exponentially small effect
on the evolution of the ground state.
In the next section we will use this phenomenon 
to obtain an exponential
lower bound on the time complexity of a perturbed version of the
Minimum Hamming Weight Problem.

\subsection{The Perturbed Hamming Weight Problem}
We will now consider the minimization of a function  
that is variation of the Hamming weight function
of the previous section:
 \begin{eqnarray} \label{eq:functionfgeneral}
f(z) & := & \left\{
\begin{array}{rl}
w(z) & \mbox{if $w(z) \leq (\frac{1}{2}+\varepsilon)n$,} \\
p(z) & \mbox{if $w(z) > (\frac{1}{2}+\varepsilon)n$,} 
\end{array}
\right.
\end{eqnarray}
with $\varepsilon>0$ and $p(z)$ a decreasing function that 
achieves the global minimum $f(z)=p(z)= -1$ in the 
$w(z) > (\frac{1}{2}+\varepsilon)n$ region. 
Our main result will be the proof that minimum gap of 
the corresponding adiabatic evolution $H_f(t)$ is
exponentially small, and hence that the 
adiabatic minimization of $f$ requires
a delay factor that is exponential in the input size $n$.

For clarity of exposition, we will focus on the special case where
\begin{eqnarray} \label{eq:functionf}
f(z) & := & \left\{
\begin{array}{rl}
w(z) & \mbox{if $z\neq 1\cdots 1$,} \\
-1 & \mbox{if $z= 1\cdots 1$.} 
\end{array}
\right.
\end{eqnarray}
The proof contains all the ingredients required for the general result
mentioned above.

The fact that this problem is a perturbed version of the 
Minimum Hamming Weight Problem is best expressed by 
\begin{eqnarray*}
H_f(t) & := & H_w(t) - \smfrac{t}{T}(n+1)\ketbra{1^n}{1^n}.
\end{eqnarray*}
We will analyze the time-dependent eigenvalues of $H_f$
by comparing them to those of $H_w$.
In the previous section, we were able
to diagonalize the $H_w$ matrix by the unitary transformation
$V(s)$ that maps the bit string 
$\ket{y_1}\ox\cdots\ox\ket{y_n}$ to the tensor
product $\ket{v_{y_1}(s)}\ox\cdots\ox\ket{v_{y_n}(s)}$.
Hence, using $s:=\smfrac{t}{T}$,
we have that $V^\dagger(s)\cdot H_w(t)\cdot V(s)$ is
a diagonal matrix with spectrum $\{\lambda_y(s)|y\in\{0,1\}^n\}$.
By looking at $H_f$ in the eigenbasis of $H_w$ 
we get the following matrix $A$, where we surpress
some of the parameters $t$ and $s$ for ease of notation:
\begin{eqnarray*} \label{eq:defA}
A & := &V^\dagger\cdot H_w\cdot V - 
s(n+1)V^\dagger\ketbra{1^n}{1^n}V.
\end{eqnarray*}
Note first that for $t=0$ and $t=T$, $A$ is a diagonal matrix.
For intermediate values of $t$, $A$ will have off-diagonal
entries caused by the perturbation 
$-s(n+1)\ketbra{1^n}{1^n}$ in the Hamiltonian $H_f$.
At $t=0$ the minimum eigenvalue is zero, which is 
indicated by the $A_{1,1}=0$ in the top-left corner 
of the Hamiltonian.  At $t=T$, the minimal eigenvalue has 
changed to $-1$ (for $z=1^n$), which coincides with
the  bottom-right element $A_{2^n,2^n}=-1$.
The eigenvectors of these values are
$\ket{v_{0}^n(0)}$ and$\ket{v_{1}^n(1)}$, 
respectively.  Intuitively, one expects the critical moment in
 the time evolution of $H_f$ to occur when the ground state has 
to change from $\ket{v_0^n}$ to $\ket{v_1^n}$.
This is indeed the case as we will see next.

To prove our  claim we will introduce another matrix $B$
that equals the matrix $A$ with its
entries $A_{2,1},\dots,A_{2^n,1}$ 
and $A_{1,2},\dots,A_{1,2^n}$ erased:
\begin{eqnarray*}
B &:=&
\left(
\begin{array}{c|ccc}
A_{1,1} & 0 & \cdots & 0 \\\hline
0 & A_{2,2}&\cdots &A_{2,2^n} \\
\vdots &\vdots & &\vdots \\
0 & A_{2^n,2}&\cdots &A_{2^n,2^n} \\
\end{array}
\right),
\end{eqnarray*}
or, equivalently,
\begin{eqnarray*}
B & := & A - A\ketbra{0^n}{0^n}
- \ketbra{0^n}{0^n}A  + 2A_{1,1}\ketbra{0^n}{0^n}.
\end{eqnarray*}
By construction, the state $\ket{v_0^n(s)}$
will be an eigenstate of $B$ for every $s$ with $A_{1,1}$
as its eigenvalue.
At $t=0$ the minimum eigenvalue of $B$
coincides with this $A_{1,1}=0$ entry; 
while at the final $t=T$ the minimum eigenvalue (with value $-1$)
is `located' in the $(2^n-1)\times(2^n-1)$ sub-matrix
(corresponding to the subspace orthogonal to $\ket{v_0^n(s)}$). 
Because $B$ transforms continuously between these
two extremes, it follows that there is a critical 
moment $s_c$ for which the minimum eigenvalue
in this subspace and the eigenvalue $A_{1,1}$
are identical. In short, at $s_c$ the matrix $B$ 
has a `zero gap' between its two minimum eigenvalues.

It can also be shown by the definitions of $A$ and $V$, 
 the fact that $V^\dagger H_w V$ is diagonal, and the
lower bound of Equation~\ref{eq:lowerbw} that:
\begin{eqnarray*}
\vnorm{A-B} & = & 
\sqrt{2}\cdot\vnorm{A\ket{0^n}-\bra{0^n}A\ketbra{0^n}{0^n}} \\
& = & 
s\sqrt{2}(n+1)\cdot|\bra{1^n}V(s)\ket{0^n}| \\
& \leq & \frac{s(n+1)}{\sqrt{2^{n-1}}}.
\end{eqnarray*}
The \emph{optimal matching distance} between $A$ and $B$ expresses
how close the spectra $\{\lambda_1,\dots,\lambda_{2^n}\}$ and 
$\{\mu_1,\dots,\mu_{2^n}\}$ of $A$ and $B$ are, 
and is formally defined by 
\begin{eqnarray*}
d(A,B)
& := & 
\min_{\pi}\max_{1\leq j \leq 2^n}|\lambda_j-\mu_{\pi(j)}|,
\end{eqnarray*}
with the minimization over all permutations $\pi\in S_{2^n}$. 
It is a known result in matrix analysis that for Hermitian 
matrices $A$ and $B$ this distance is upper bounded 
by $\vnorm{A-B}$  (see Section~VI.3 in \cite{bhatia}).

We thus reach the conclusion that for all values of $s$,
the gap $g(s)$ of $A$ (and hence of $H_f(s)$) will never 
be bigger than the gap of $B$ plus twice the distance $\vnorm{A-B}$.
At the critical moment $s_c$, when the two minimal eigenvalues of $B$ 
are identical, this implies for the gap of $A$ the upper bound 
$g(s_c) \leq  {s_c(n+1)}/{\sqrt{2^{n-3}}}$, 
and hence also for the Hamiltonian $H_f$: 
$g_\mathrm{min} \in O(\smfrac{n}{\sqrt{2^n}})$.
Applying the requirement $T\gg g_\mathrm{min}^{-2}$ thus 
yields the lower bound $\Omega(\smfrac{2^n}{n^2})$ for the 
delay factor $T$.

\subsection{Generalization}
It is not difficult to see that the above lower bound method 
applies to the larger class of functions mentioned 
in Equation~\ref{eq:functionfgeneral}. 
The critical property of $f$ is that it only deviates 
from the Hamming weight function $w(z)$ for those strings $z$ 
that have an exponential small inner-product with the 
$H(s)$ ground state $\ket{v_0(s)\cdots v_0(s)}$
for all $s$ 
(the property of Equation~\ref{eq:lowerbw}).

As long as the perturbation $p:\{0,1\}^n\rightarrow \R$ in 
Equation~\ref{eq:functionfgeneral} is polynomial in $n$, 
we have an inequality similar to Equation~\ref{eq:lowerbw}:
\begin{eqnarray}\label{eq:lowerbp}
\vnorm{(H_f-H_w)\ket{v_0(s)\cdots v_0(s)}} & \in &
2^{-\Omega(n)}.
\end{eqnarray}
Hence, if the perturbation $p$ is such that the minimum of 
$f$ is not $f(0^n)$, then the adiabatic algorithm requires
a delay $T \gg g^{-2}_{\mathrm{min}}$ that is exponential in the 
input size of the problem. i.e. $T \in 2^{\Omega(n)}$.

\section{Conclusions}
Adiabatic quantum computation is a novel paradigm for the design of
quantum algorithms --- it is truly quantum in the sense that it can
be used to speed up searching by a quadratic factor over any classical
algorithm. On the question of whether this new paradigm may be used
to efficiently solve NP-complete problems on a quantum computer ---
we showed that the usual query complexity arguments cannot be
used to rule out a polynomial time solution. On the other hand, we
argue that the adiabatic approach may be thought of as a kind of
`quantum local search'. We designed a family of minimization
problems that is hard for such local search heuristics, 
and established an 
exponential lower bound for the adiabatic algorithm for these problems. 
This provides insights into the limitations of this approach. In an upcoming
paper~\cite{dv}, we generalize these techniques to show a similar
exponential slowdown for 3SAT. 
It remains an open question whether adiabatic quantum computation 
can establish an exponential speed-up over traditional computing 
or if there exists a classical algorithm that can simulate the
quantum adiabatic process efficiently.

\section{Acknowledgements}
We wish to thank Dorit Aharonov and Tad Hogg for many useful 
discussions.

 \appendix
\section{Proof of the Query Complexity Result}
\setcounter{theorem}{1}
\begin{theorem}
Let $b \in \{0,1\}^n$ and let $I$ be the subset of
$\{1,\dots, n\}$ such that $b_i = 1 \iff i \in I$. Then,
\begin{eqnarray*}
F(b) &=& F(0^n) + \sum_{i \in I}{Y_i} +
\sum_{i < j\in I}{Y_{ij}} + \sum_{i<j<k \in I}{Y_{ijk}}.
\end{eqnarray*}
Furthermore,
\begin{eqnarray*}
 Y_i &=& F(e^i) - F(0^n) \\
Y_{ij}& =& F(e^{ij}) - F(e^i) - F(e^j) +F(0^n) \\
 Y_{ijk} &=& F(e^{ijk}) - F(e^{ij}) - F(e^{ik})-F(e^{jk})
  + F(e^i) + F(e^j) + F(e^k) -F(0^n).
\end{eqnarray*}
\end{theorem}
\begin{proof}
We count the total number of unsatisfied clauses by analyzing each
type of clause.

Firstly, the only clauses of the form
$\overline{X}\overline{X}\overline{X}$ that will not be satisfied
are those that have all three variables with indices in $I$. This
gives us
\begin{eqnarray*}
\sum_{i < j < k \in I}{|\overline{X_i}\overline{X_j}
\overline{X_k}|}
\end{eqnarray*}
unsatisfied clauses of the form
$\overline{X}\overline{X}\overline{X}$. Note that if there are
less than $3$ ones in $b$ then any of the summations over $i,j,k
\in I$ satisfying $i < j < k$ will be empty and thus sum to $0$.

Secondly, the only clauses of the form $\overline{X}\overline{X}
X$ that will not be satisfied are those that have both of the
negated variables with indices in $I$ and the positive variable
with index not in $I$. This gives us
\begin{eqnarray*}
& &  \sum_{i< j \in I}{|\overline{X_i}\overline{X_j}X|}
 - \sum_{i < j < k \in
I}{\left(|\overline{X_i}\overline{X_j} X_k| +
|\overline{X_i}\overline{X_k}X_j| + |\overline{X_j}\overline{X_k}
X_i|\right)}
\end{eqnarray*}
unsatisfied clauses of the form $\overline{X}\overline{X}X$.

Thirdly, the only clauses of the form $\overline{X} X X$ that will
not be satisfied will be those that have the negated variable with
index in $I$ and the positive variables with indices not in $I$.
This gives us
\begin{eqnarray*}
& & \sum_{i \in I}{|\overline{X_i}XX|} - 
\sum_{i<j \in I}{\left(|\overline{X_i} X_j X| + |\overline{X_j}X_i X|\right)}
+ \sum_{i<j<k \in I}{\left(|\overline{X_i}X_j X_k| + |
\overline{X_j}X_i X_k | + | \overline{X_k}X_i X_j |\right)}
\end{eqnarray*}
unsatisfied clauses of the form $\overline{X} X X$.

The only clauses of the form $XXX$ that will not be satisfied are
those that contain no variable with index in $I$. This gives us
\begin{eqnarray*}
& & 
|XXX| - \sum_{i \in I}{|X_i XX|} + \sum_{i < j \in I}{|X_i X_j X|}  
- \sum_{i < j <k \in I}{|X_i X_j X_k|}
\end{eqnarray*} 
unsatisfied clauses of the form $XXX$.

Similarly, we have
\begin{eqnarray*} & &  \sum_{i<j \in I}{|
    \overline{X_i}\overline{X_j}|} 
\end{eqnarray*}
unsatisfied clauses of the form $\overline{X}\overline{X}$,
\begin{eqnarray*} & &  \sum_{i \in I}{|\overline{X_i}X|} -
\sum_{i <j \in I}{\left(|\overline{X_i}X_j| +
|\overline{X_j}X_i|\right)} 
\end{eqnarray*}
 unsatisfied clauses of the form
$\overline{X}X$,
\begin{eqnarray*} & &  |XX| - \sum_{i \in I}{|X_iX|} + 
\sum_{i<j \in I}{|X_i X_j|} 
\end{eqnarray*}
unsatisfied clauses of the form $XX$,
\begin{eqnarray*} & &  \sum_{i \in I}{|\overline{X_i}|} 
\end{eqnarray*}
unsatisfied clauses of the form $\overline{X}$, and
\begin{eqnarray*} & &  |X| - \sum_{i \in I}{|X_i|} 
\end{eqnarray*}
unsatisfied clauses of the form $X$.

These account for all the unsatisfied clauses.  Summing these
quantities while rearranging terms according to the number of
variables in the summations, gives us the first part of the
theorem:
\begin{eqnarray*}
F(b)& =& |XXX| + |XX| + |X| 
 + \sum_{i \in I}{Y_i} +
\sum_{i < j \in I}{Y_{ij}}  
+ \sum_{i<j<k \in I}{Y_{ijk}} \\
& = & F(0^n) + \sum_{i \in I}{Y_i} + \sum_{i < j \in I}{Y_{ij}} +
\sum_{i<j<k \in I}{Y_{ijk}}.
\end{eqnarray*}

Notice that for $F(e^i)$ any of the summations with more
than one variable will be empty, and we get
\begin{eqnarray*}
F(e^i) &=& F(0^n) + Y_i.
\end{eqnarray*}
Similarly, for $F(e^{ij})$ any of the summations with
three variables will be empty, and we are left with
\begin{eqnarray*}
F(e^{ij}) &=& F(0^n) + Y_i + Y_j + Y_{ij} .
\end{eqnarray*}
Lastly, for $F(e^{ijk})$ we get
\begin{eqnarray*}
F(e^{ijk})
&=& F(0^n) + Y_i + Y_j + Y_k + Y_{ij} + Y_{ik} + Y_{jk} 
 + Y_{ijk}.
\end{eqnarray*}
From these equations follow the second part of the theorem.
\end{proof}

\begin{thebibliography}{99}

\bibitem{AIV}
S.\ Arora, R.\ Impagliazzo, and U.\ Vazirani,
``On the Role of the Cook-Levin Theorem in Complexity Theory'',
manuscript (1994)

\bibitem{bbbv}
C.\ Bennett, E.\ Bernstein, G.\ Brassard, and U.\ Vazirani,
``Strengths and weaknesses of quantum computing'',
\emph{SIAM Journal on Computing,} Volume~26, No.~5, pages~1510--1523 (1997);
quant-ph archive, report no.~9701001

\bibitem{bhatia}
R.\ Bhatia, \emph{Matrix Analysis,} Graduate Texts in Mathematics, 
Volume 169, Springer-Verlag (1997)

\bibitem{bornfock}
M.\ Born and V.\ Fock, 
``Beweis des Adiabatensatzes'',
\emph{Zeitschrift f{\"u}r Physik,} Volume 51, 
pp.~165--180 (1928)
 
\bibitem{dv}
W.\ van Dam and U.\ Vazirani,
``On the Power of Adiabatic Quantum Computation'',
in preparation

\bibitem{farhi00}
E.\ Farhi, J.\ Goldstone, S.\ Gutmann, and M.\ Sipser,
``Quantum Computation by Adiabatic Evolution'', 
quant-ph report no.~0001106 (2000)

\bibitem{farhi01}
E.\ Farhi, J.\ Goldstone, S.\ Gutmann, J.\ Lapan, A.\ Lundgren, and D.\ Preda, 
``A Quantum Adiabatic Evolution Algorithm Applied to Random Instances of an
NP-Complete Problem'', \emph{Science,} Volume~292, April, pp.~472--476 (2001) 

\bibitem{farhi3sat}
E.\ Farhi, J.\ Goldstone, S.\ Gutmann
``A Numerical Study of the Performance of a Quantum Adiabatic 
Evolution Algorithm for Satisfiability'', 
quant-ph report no.~0007071 (2000)

\bibitem{farhisearch}
E.\ Farhi and S.\ Gutmann,
``Analog analogue of digital quantum computation'',
\emph{Physical Review A,} Volume~57, Number 4, pp.~2403--2406 (1998);
quant-ph report no.~9612026 

\bibitem{grover}
L.\ Grover,
``A fast quantum mechanical algorithm for database search'',
\emph{Proceedings of the 28th Annual ACM Symposium on the 
Theory of Computing,}
pp.~212--219 (1996);
quant-ph report no.~9605043

\bibitem{hogg}
 T.\ Hogg,
 ``Quantum Search Heuristics'',
 \emph{Physical Review A,} Volume~61, Issue~5, 
 pp.~052311 (2000)

\bibitem{messiah}
A.\ Messiah, \emph{Quantum Mechanics,}
John Wiley \& Sons (1958)

\bibitem{shor}
P.\ Shor.
``Algorithms for quantum computation: Discrete logarithms and
  factoring'',
\emph{SIAM Journal on Computing,} Volume~26, No.~5, pp.~1484--1509 (1997);
 quant-ph report no.~9508027

\bibitem{Va}
U.\ Vazirani.
``On the Power of Quantum computation'',
\emph{Philosophical Transactions of the Royal Society of London A,} 
Volume~356, Number~1743, pp.~1759--1768 (1998)
\end{thebibliography}
\end{document}